\documentclass[a4paper,11pt]{article}
\pdfoutput=1 % if your are submitting a pdflatex (i.e. if you have
\usepackage{jinstpub} % for details on the use of the package, please
\usepackage{subfigure}

\title{Investigation of thin n-in-p planar pixel modules for the ATLAS upgrade}

\author[1]{N. Savic,\note{Corresponding author.}}
\author[]{J. Beyer,}
\author[]{A. La Rosa,}
\author[]{A. Macchiolo,}
\author[]{R. Nisius}

\affiliation[a]{Max-Planck-Institut f\"ur Physik (Werner-Heisenberg-Institut),\\ F\"ohringer Ring 6, D-80805 M\"unchen, Germany}

\emailAdd{Natascha.Savic@mpp.mpg.de}

\abstract{In view of the High Luminosity upgrade of the Large Hadron Collider (HL-LHC), planned to start around 2023-2025, the ATLAS experiment will undergo a replacement of the Inner Detector. A higher luminosity will imply higher irradiation levels and hence will demand more radiation hardness especially in the inner layers of the pixel system. The n-in-p silicon technology is a promising candidate to instrument this region, also thanks to its cost-effectiveness because it only requires a single sided processing in contrast to the n-in-n pixel technology presently employed in the LHC experiments. In addition, thin sensors were found to ensure radiation hardness at high fluences. An overview is given of recent results obtained with not irradiated and irradiated n-in-p planar pixel modules. The focus will be on  n-in-p planar pixel sensors with an active thickness of 100 and 150~$\mathrm{\mu}$m recently produced at ADVACAM. To maximize the active area of the sensors, slim and active edges are implemented. The performance of these modules is investigated at beam tests and the results on edge efficiency will be shown.}


\proceeding{18$^{\text{th}}$ International Workshop on Radiation Imaging Detectors\\
	3-7 July 2016\\
	Barcelona, Spain}

\begin{document}
	\maketitle
	\flushbottom
	
	\section{ATLAS pixel upgrade for HL-LHC}
		At HL-LHC the ATLAS pixel system is expected to be exposed to fluences up to 2$\times$10$^{16}$ $\mathrm{n}_{\mathrm{eq}}/\mathrm{cm}^2$ (1 MeV neutron equivalent) \cite{atlasUp}. For this, planar pixel modules based on the n-in-p technology and employing thin sensors have been developed. The modules consist of 100 and 150~$\mathrm{\mu}$m thick sensors produced at ADVACAM (Finland) and CiS (Germany). At ADVACAM even thinner sensors with a thickness of 50~$\mathrm{\mu}$m were produced. In addition, the active edge technology was implemented to allow for a reduction of the dead area at the periphery of the devices since an overlap of pixel modules along the beam direction is not possible due to space limitations in the innermost region. The sensors are interconnected with solder bump-bonding to ATLAS FE-I4 read-out chips and characterized with particles from radioactive sources and test beams at the CERN-SpS and DESY II. The results of beam test measurements are discussed for unirradiated modules with a sensor thickness of 100~$\mathrm{\mu}$m. A module from a production of VTT (Finland) with the same sensor thickness was irradiated to a fluence of 1$\times$10$^{16}$ $\mathrm{n}_{\mathrm{eq}}/\mathrm{cm}^2$. Results on the hit efficiency and power dissipation reveal that it is a promising candidate for the innermost layer. To manage the forthcoming high occupancy at HL-LHC, smaller pixel dimensions with respect to the ones currently implemented in the FE-I3 chip (50x400~$\mathrm{\mu}$m$^{2}$) \cite{fei3} and the FE-I4 chip (50x250~$\mathrm{\mu}$m$^{2}$) \cite{fei4} are necessary. The new read-out chip for the future ATLAS pixel systems with a pixel cell of 50x50~$\mathrm{\mu}$m$^{2}$ produced in the 65 nm CMOS technology is being developed by the CERN RD53 Collaboration \cite{rd53} and the first prototype, the RD53A chip, is foreseen to be ready in 2017 \cite{chip}. Highly segmented sensors will represent a challenge for the tracking in the forward region of the pixel system at HL-LHC. At CiS sensors have been already produced with 50x50~$\mathrm{\mu}$m$^{2}$ pixel implants, connected by a metal layer in such a way to be still compatible with the FE-I4 chip footprint. First results on the in-pixel charge collection and efficiency of small pixel implants before irradiation will be shown at perpendicular beam incidence. To reproduce the performance of 50x50~$\mathrm{\mu}$m$^{2}$ pixels at high pseudo-rapidity, FE-I4 sensors of different thicknesses have been studied in beam tests at high incidence angle (80$^{\circ}$) with respect to the direction of the short pixel dimension. Cluster shapes, charge collection and hit efficiency at different threshold values are investigated and compared.
	
	\section{Pixel module characterization}
	Two different experimental setups were used to investigate the performance of pixel modules. Charge collection measurements have been carried out with $^{90}$Sr $\beta$-electrons using the USBPix and RCE readout systems \cite{usbpix, rce}. The charge calibration was optimized utilizing $^{241}$Am and $^{109}$Cd $\gamma$-sources as reference. The modules are operated inside a climate chamber at a stable ambient temperature of 20$^{\circ}$C before irradiation. To determine the hit efficiency beam test measurements have been performed at the DESY II in Hamburg with 5 GeV electrons and at the SpS at CERN with 120 GeV pions with the EUDET telescope. This beam telescope provides a precise reference track trajectory using its 3 planes of fine-pitch MIMOSA26 sensors upstream  the Device Under Test (DUT) and another 3 planes downstream the DUTs. Each EUDET plane can reach 3.5~$\mathrm{\mu}$m of spatial resolution in case of low energetic electrons and 2~$\mathrm{\mu}$m for high energetic pions \cite{hendrick, jens}. The hit efficiency of the DUTs is defined as the ratio of the number of reconstructed tracks with matching hits in the DUT to the total number of reconstructed tracks passing through the DUT determined by performing track reconstruction with the EUTelescope software \cite{eutelescope}. An absolute systematic uncertainty of 0.3\% is associated to all hit efficiency measurements as estimated in Ref.~\cite{jens}.
	
	\subsection{The thin n-in-p pixel production with active and slim edges at ADVACAM}
\begin{figure}[t!]
	\centering     %%% not \center
	\subfigure[]{\label{fig:a}\includegraphics[width=70mm, height=47mm]{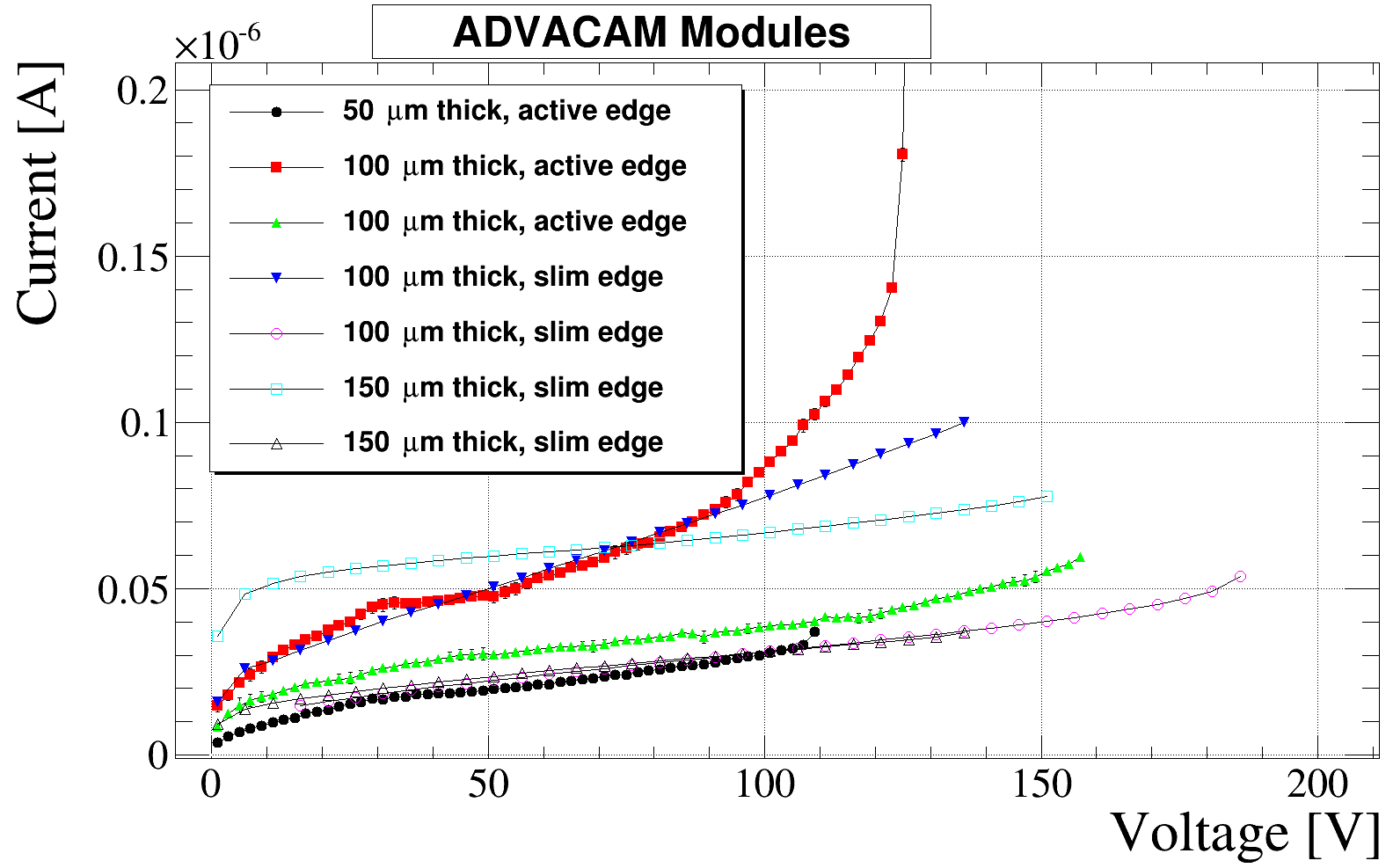}}
	\subfigure[]{\label{fig:b}\includegraphics[width=70mm]{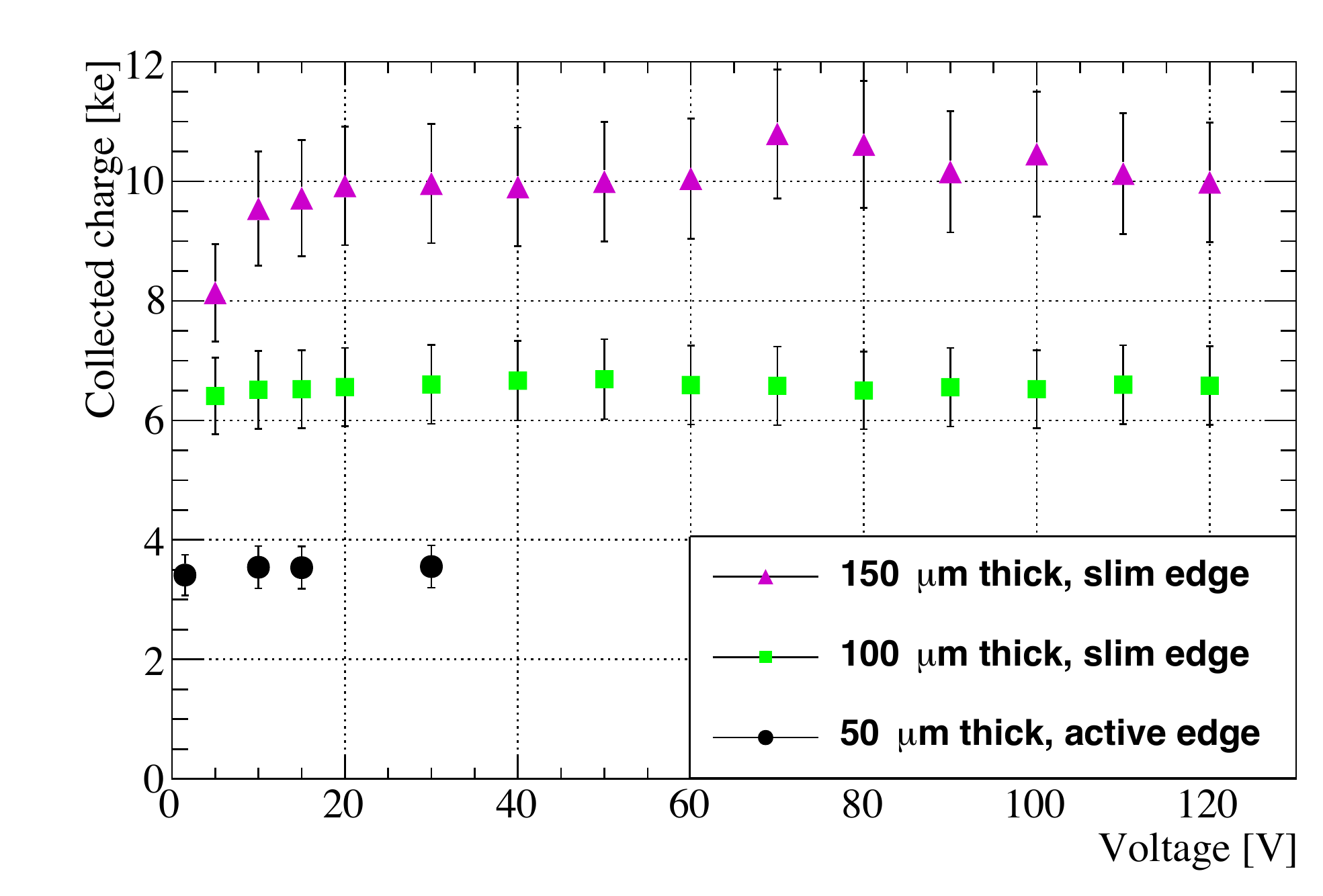}}
	\caption{Properties of FE-I4 pixel modules from the ADVACAM production with sensors of 50, 100 and 150~$\mathrm{\mu}$m thickness. The leakage current as a function of bias voltage is shown in (a) and the collected charge as a function of bias voltage, measured in $^{90}$Sr source scans, is displayed in (b).}
	\label{adv}
\end{figure}

\begin{figure}[t!]
	\centering     %%% not \center
	\subfigure[Module with active edge sensor]{\label{fig:a}\includegraphics[width=65mm]{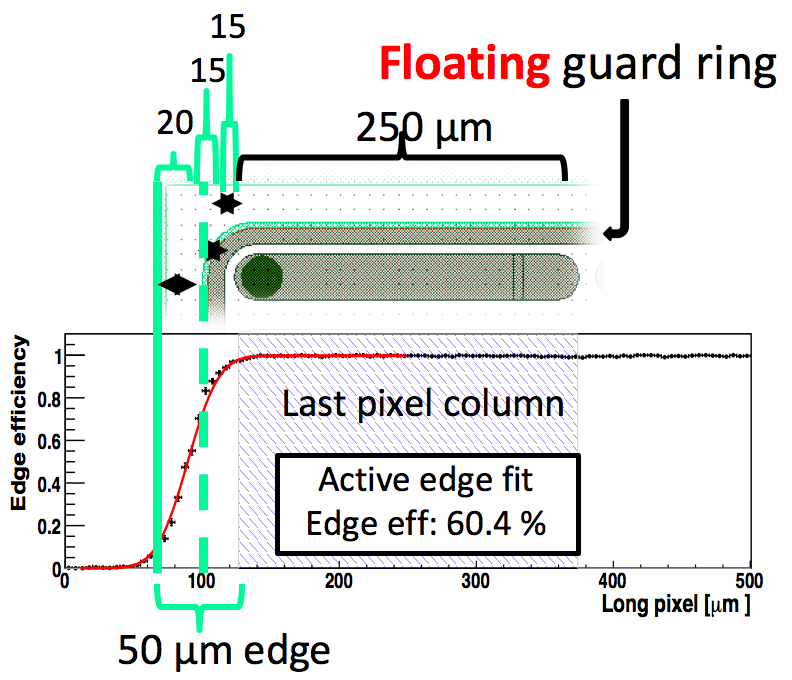}}
	\subfigure[Module with slim edge sensor]{\label{fig:b}\includegraphics[width=65mm]{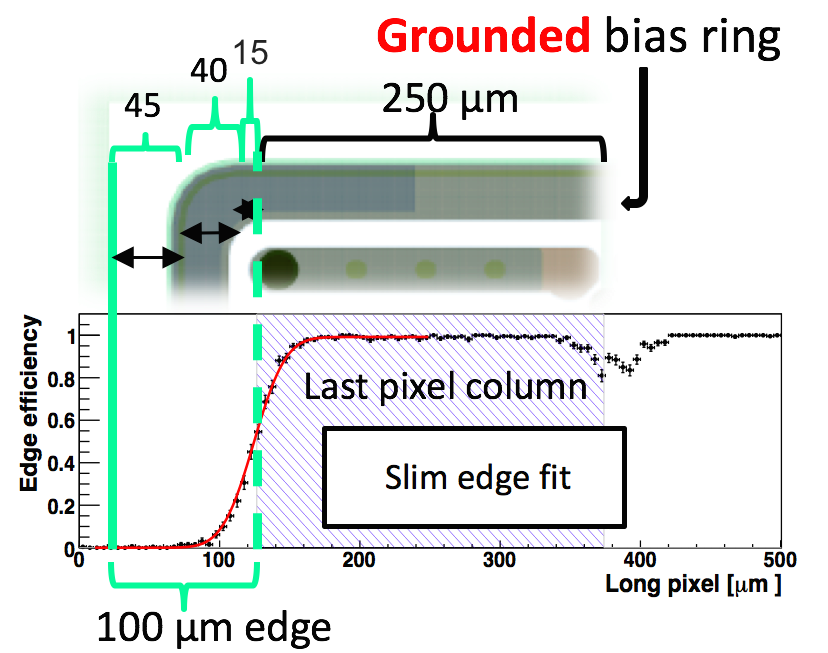}}
	\caption{Hit efficiency at the edge of the 100~$\mathrm{\mu}$m thin sensor with 50~$\mathrm{\mu}$m active edge (a) and 100~$\mathrm{\mu}$m slim edge (b) design before irradiation. The pixel cell has a standard size of 50x250~$\mathrm{\mu}$m$^{2}$ and the hit efficiency is evaluated as a function of the distance from the last pixel column to the edge region. Smearing of the position of approximately 18~$\mathrm{\mu}$m is due to the telescope resolution.}
	\label{edge}
\end{figure}

	The second Silicon on Insulator (SOI) production at ADVACAM employing 6'' Float Zone (FZ) wafers is focused on planar n-in-p pixel sensors with active edges. It has been designed to extend the research program initiated with the VTT FE-I3 compatible sensors (see \cite{phil}) to FE-I4 geometries in the thickness range of 50 to 150~$\mathrm{\mu}$m. Trenches are realized around the perimeter of each sensor with Deep Reactive Ion Etching (DRIE) using the mechanical support offered by the handle wafer of the SOI stack. The boron implantation at the backside of the p-type sensor is then extended to the edges with a four-quadrant ion implantation \cite{edge}. Subsequently, the handle wafer is removed and homogeneous p-spray is used for the inter-pixel isolation. A slim edge and an active edge design have been implemented differentiating by the distance d$_{e}$ of the last pixel implant and the sensor edge. Two variants of slim edge sensors with d$_{e}$=100~$\mathrm{\mu}$m are produced: the first one with one bias ring (BR), the second with one BR together with one floating guard ring (GR). In both designs a standard punch-through structure has been implement as well as a common punch-through structure in the design of a single BR. More detailed explanations and investigations on punch-through mechanism and structure are given in \cite{nani}. The active edge design characterized by d$_{e}$= 50~$\mathrm{\mu}$m includes only one floating GR. No punch-through structure is implemented in this design. Results of a first electrical characterization of all sensor types are summarized in Fig.~\ref{adv} and show a good performance. The breakdown voltages in Fig.~\ref{adv}(a) of around 150 V are well above the thickness dependent depletion voltages of 10 to 25 V. In Fig.~\ref{adv}(b) the collected charge is shown as a function of bias voltage for active and slim edge designs for all sensor thicknesses. The collected charge is flat for all sensor thicknesses after depletion and the measured values can be well accounted for by the expectation (see \cite{phil}): for 50~$\mathrm{\mu}$m a charge of 3.1~ke was collected, for 100~$\mathrm{\mu}$m 6.7~ke and for 150~$\mathrm{\mu}$m 10.2~ke. An analysis of the performance at the sensor edge has been carried out for these devices with data from a beam test with 5 GeV electrons at DESY. The slim edge design with one BR and the common punch-through structure and the active edge design with one GR and no punch-through structure, both for a sensor thickness of 100~$\mathrm{\mu}$m, are displayed in Fig.~\ref{edge}. For the active edge sensor an efficiency of 99.8\% within the last pixel column and an average efficiency of 60.4\% up to the sensor edge was achieved before irradiation. The slim edge sensor yields an efficiency of 99.2\% in the last pixel column.

	\subsubsection{Performance at high incident angle}
	\subsubsection{Performance at high incident angle}
	Smaller pixel cells are challenging for the tracking in regions of high pseudo-rapidity (high $\eta$). To make predictions on the performance of a 50x50~$\mathrm{\mu}$m$^{2}$ pixel cell at $\eta$=2.5 FE-I4 modules were arranged such that the particles of the beams at DESY and CERN were crossing the pixel along the short side (50~$\mathrm{\mu}$m) at a nominal angle of $\vartheta$=80$^{\circ}$.
	\begin{figure}[t]
		\centering     %%% not \center
		\subfigure[]{\label{cluster}\includegraphics[width=70mm]{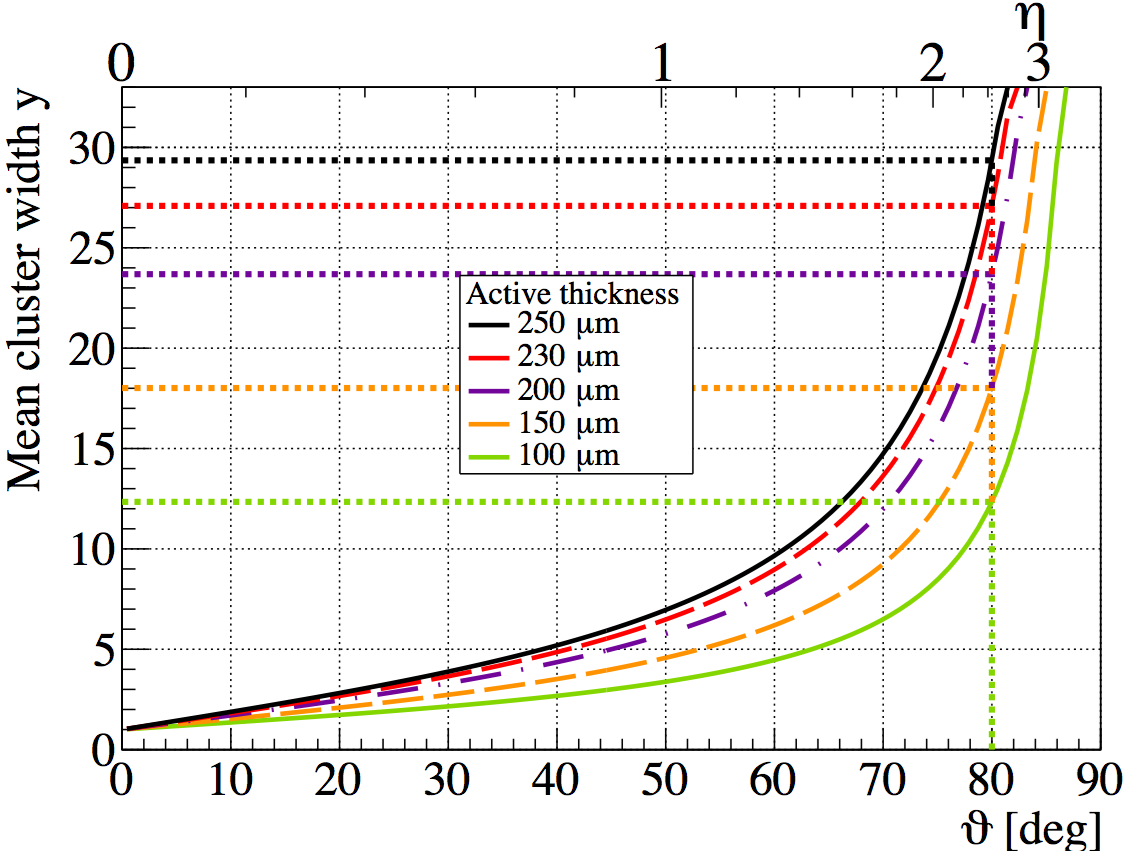}}
		\subfigure[]{\label{highphichargeallthick}\includegraphics[width=70mm]{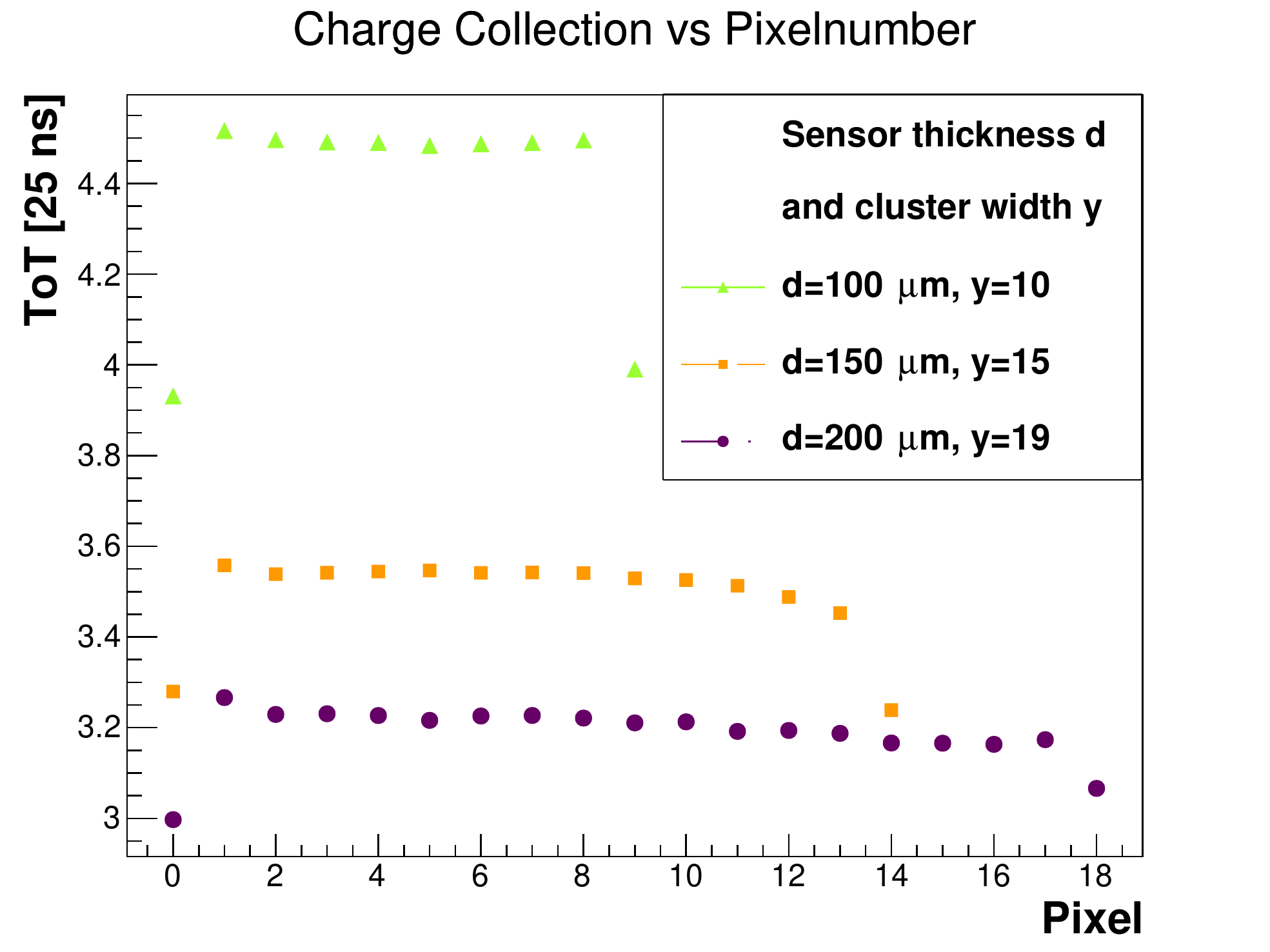}}
		\caption{(a) Expected mean cluster width along the short pixel cell direction for an FE-I4 module as a function of beam incidence angle. (b) Charge collection in units of Time over Threshold (ToT), as a function of pixel number for different cluster sizes obtained with unirradiated FE-I4 modules employing sensor thicknesses of 100, 150 and 200~$\mathrm{\mu}$m. All modules were tuned to a threshold of 1000 e. Pixel number zero always corresponds to the sensor front-side and is characterized by a lower charge since it is only partially traversed.}
		\label{highphidiffthick}
	\end{figure}
	\begin{figure}[htpb!]
		\centering     %%% not \center
		%\subfigure[]{\label{cluster}\includegraphics[width=70mm]{ToT_unirr_600to1000thresh_compare_corrected.pdf}}
		\includegraphics[width=70mm]{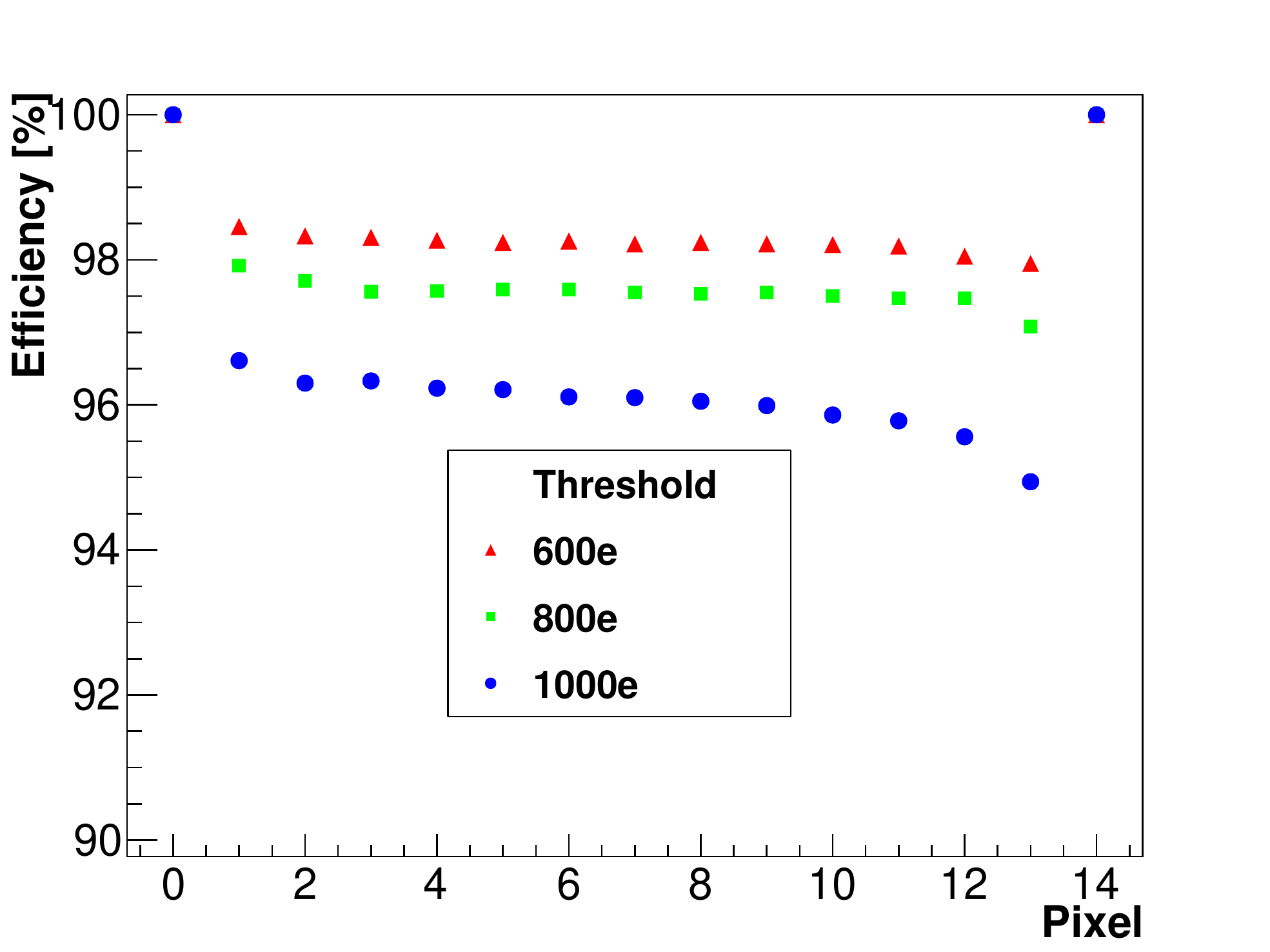}
		\caption{Hit efficiency as a function of pixel number at different target thresholds of 1000, 800 and 600~e for a not irradiated FE-I4 module with a sensor thicknesses of 150~$\mathrm{\mu}$m tilted by 80$^{\circ}$ with respect to the beam direction. The efficiencies of the first and the pixels are by construction 100 \% since they define the start and the end of the cluster.}
		\label{highphidiffthresh}
	\end{figure}
	The performance of an ADVACAM module with a sensor thickness of 150~$\mathrm{\mu}$m is compared to previous results of modules with sensor thicknesses of 100 and 200~$\mathrm{\mu}$m before irradiation \cite{nani}. The expected cluster width along $\eta$ strongly depends on the sensor thickness as visualised in Fig.~\ref{highphidiffthick}(a) and thinner sensors produce smaller clusters \cite{anna, nani}. This is in good agreement with the observed cluster widths for the sensor thicknesses of 100, 150 and 200~$\mathrm{\mu}$m displayed in Fig.~\ref{highphidiffthick}(b). An average cluster width of 15 pixels was reconstructed for the module with a 150~$\mathrm{\mu}$m thick sensor, slightly less than the expected value of 17-18. This difference was traced back to a misplacement of 2-3$^{\circ}$ of the modules with respect to the nominal position. \par
	For precise measurements of the collected charge a dedicated tuning was performed with a low target threshold of 1~ke. The average collected charge increased from 3.1~ke for a 200~$\mathrm{\mu}$m thick sensor to 3.5~ke for a 150~$\mathrm{\mu}$m and further to 4.5~ke for a sensor with a thickness of 100~$\mathrm{\mu}$m. The increase could be explained with a reduced effect of charge sharing due to a smaller lateral diffusion of charge carriers in thinner sensors. Furthermore, the ADVACAM module with a 150~$\mathrm{\mu}$m thick sensor was tuned to even lower target thresholds of 800 and 600 e, in the range considered routinely achievable by the future RD53A read-out chip. By lowering the threshold to 600 e instead of 1~ke the efficiency increased by 2\% to 98\%. The hit efficiency at different target thresholds is shown in Fig.~\ref{highphidiffthresh}.
	
	\subsection{The thin n-in-p pixel production at CiS}
	A production of thin planar n-in-p pixel on 4'' wafers on p-type FZ silicon has been recently completed at CiS. The wafers were locally, i.e. below each single sensor, thinned to an active thickness of 100 and 150~$\mathrm{\mu}$m by using anisotropic KOH etching creating backside cavities in the wafer, with a starting thickness around 500~$\mathrm{\mu}$m, leaving frames around single structures. This technology does not need the use of any handle wafer during the thinning process, thus being potentially cheaper than the alternative methods employing SOI wafers. A more detailed description of the production process is given in Ref.~\cite{nani}. Two pixel designs were implemented together with the common punch-through design which was found to retain the highest efficiencies after irradiation \cite{nani}: the standard FE-I4 pixel pitch with a 50x250~$\mathrm{\mu}$m$^{2}$ cell and a modified 50x250~$\mathrm{\mu}$m$^{2}$ pixel cell composed by five 50x50~$\mathrm{\mu}$m$^{2}$ pixel implants connected by metal lines allowing for readout by the FE-I4 chip. 
\begin{figure}[htpb!]
	\centering
	\includegraphics[width=0.47\textwidth]{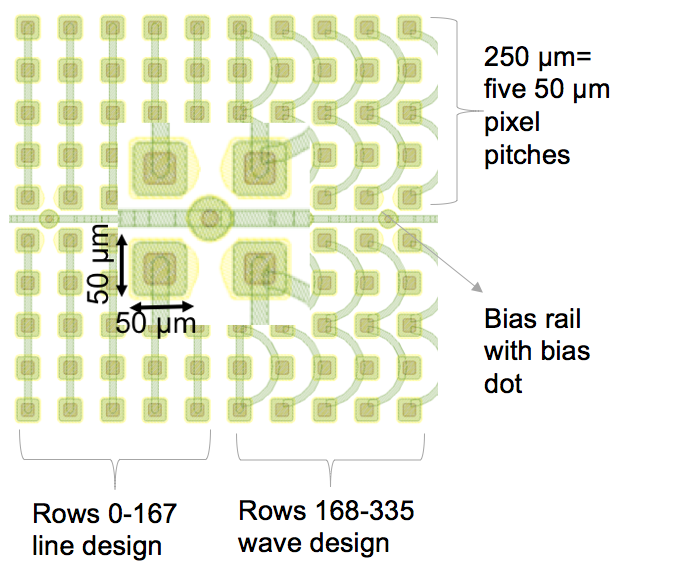}
	\caption{CiS sensor layout with a modified 50x250~$\mathrm{\mu}$m$^{2}$ pixel cell including five 50x50~$\mathrm{\mu}$m$^{2}$ pixel implants. Two different arrangements of the metal rail linking the small implants to a 50x250~$\mathrm{\mu}$m$^{2}$ pixel cell are implemented in one sensor.}
	\label{Bild}
\end{figure}
	A cut-out of the sensor layout is shown in Fig.~\ref{Bild}. Two layouts to link five 50x50~$\mathrm{\mu}$m$^{2}$ implants using a metal rail were designed and implemented in one single sensor of 336 rows and 80 columns allowing for a direct comparison of the performances. In the first 168 rows of the sensor adjacent implants were linked recreating the standard FE-I4 geometry and in the following this is referred to as line design. In constrast, in the last 168 rows neighbouring pixel implants are read out by two different channels, with only every second pixel implant inside a 50x250~$\mathrm{\mu}$m$^{2}$ being read-out by the same channel, denoted by wave design. The latter design makes it possible to investigate the effect of charge sharing in adjacent 50x50~$\mathrm{\mu}$m$^{2}$ pixels. 
\begin{figure}[tbph!]
	\centering     %%% not \center
	\subfigure[line design]{\label{effvtt}\includegraphics[width=70mm]{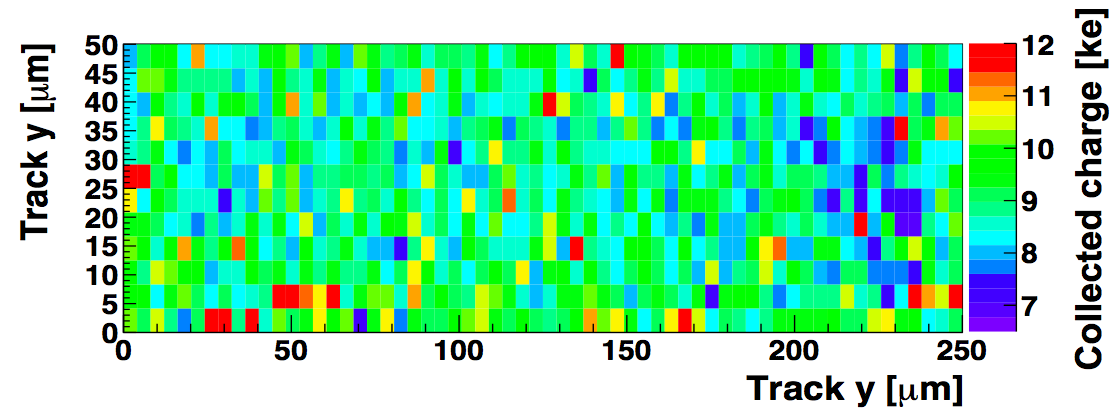}}
	\subfigure[wave design]{\label{powerdiss}\includegraphics[width=70mm]{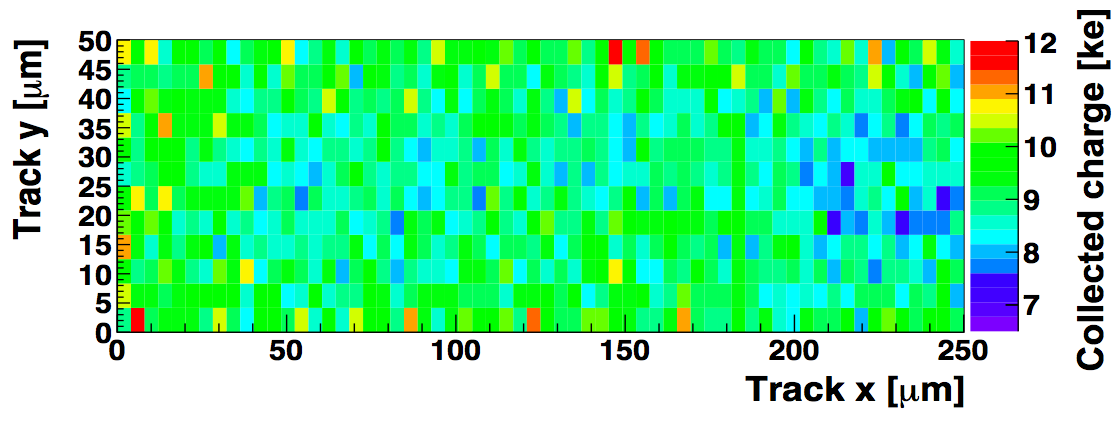}}
	\subfigure[line design]{\label{effvtt}\includegraphics[width=70mm]{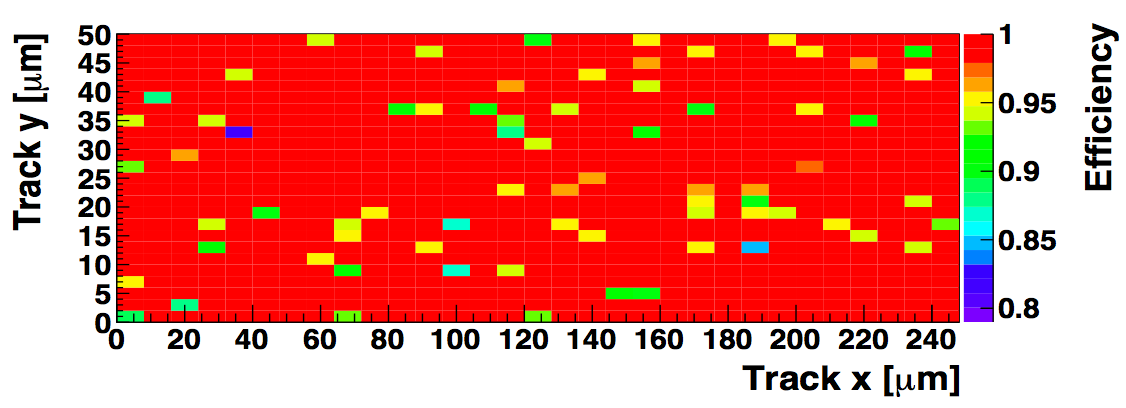}}
	\subfigure[wave design]{\label{powerdiss}\includegraphics[width=70mm]{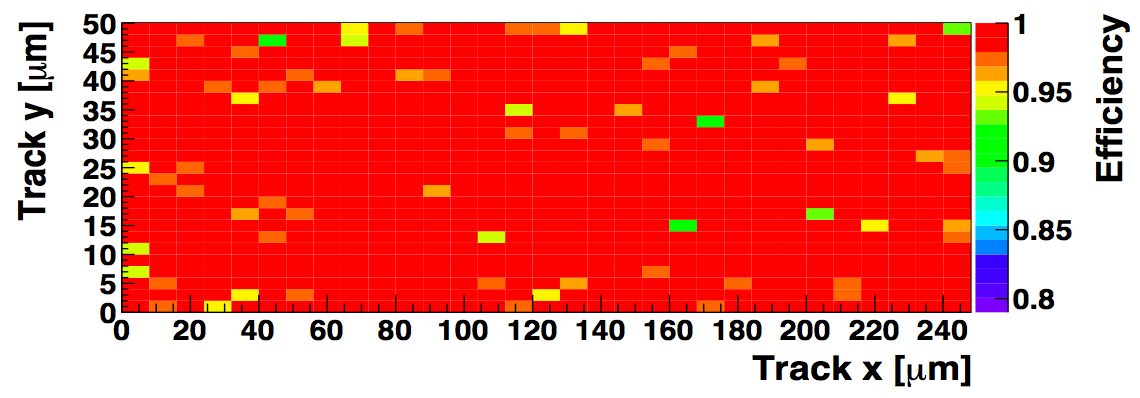}}
	\caption{Collected charge and hit efficiency within the pixel cell for the line (a, c) and wave design (b, d). The module with a 150~$\mathrm{\mu}$m thick sensor was operated at 75 V before irradiation.}
	\label{cis4results}
\end{figure}
	A module with a sensor thickness of 150~$\mathrm{\mu}$m was tested in a beam test at CERN. Both designs show good performance before irradiation with compatible values of collected charge around 10~ke and hit efficiencies of 99.3\% (line design) and 99.6\% (wave design) yielding the same values within uncertainties (see Fig.~\ref{cis4results}). This can be explained by the high charge deposit above the threshold before irradiation. Although the charge is splitted between two adjacent pixel readout channels in the wave design, both fractions of the charge are still above threshold not affecting the efficiency. Hence, there is no significant effect of the charge sharing among neighbouring pixels before irradiation.
	
	\subsection{Performance of a highly irradiated VTT module of 100~$\mathrm{\mu}$m sensor thickness}
\begin{figure}[t!]
		\centering     %%% not \center
		\subfigure[]{\label{effvtt}\includegraphics[width=70mm]{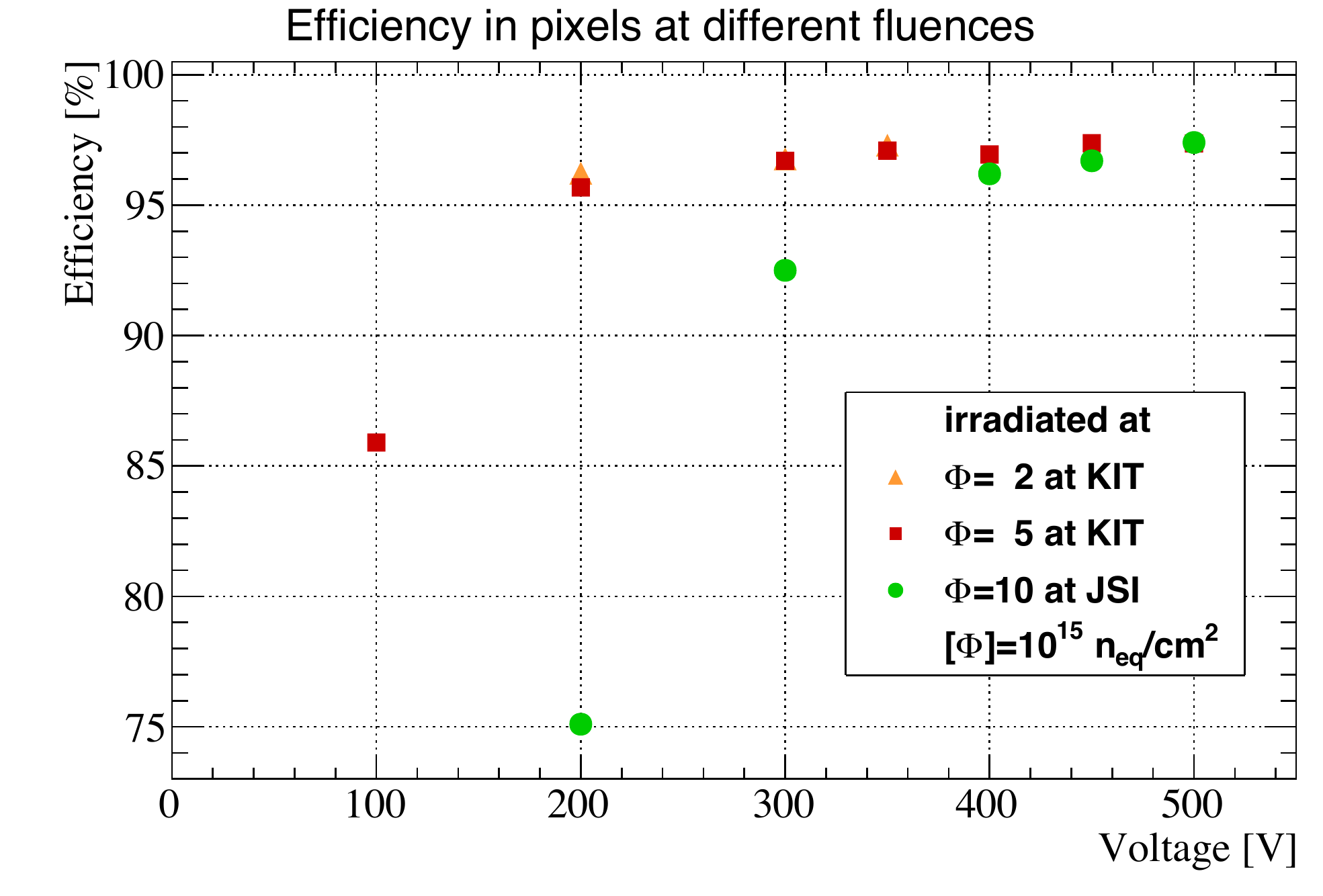}}
		\subfigure[]{\label{powerdiss}\includegraphics[width=70mm, height=50mm]{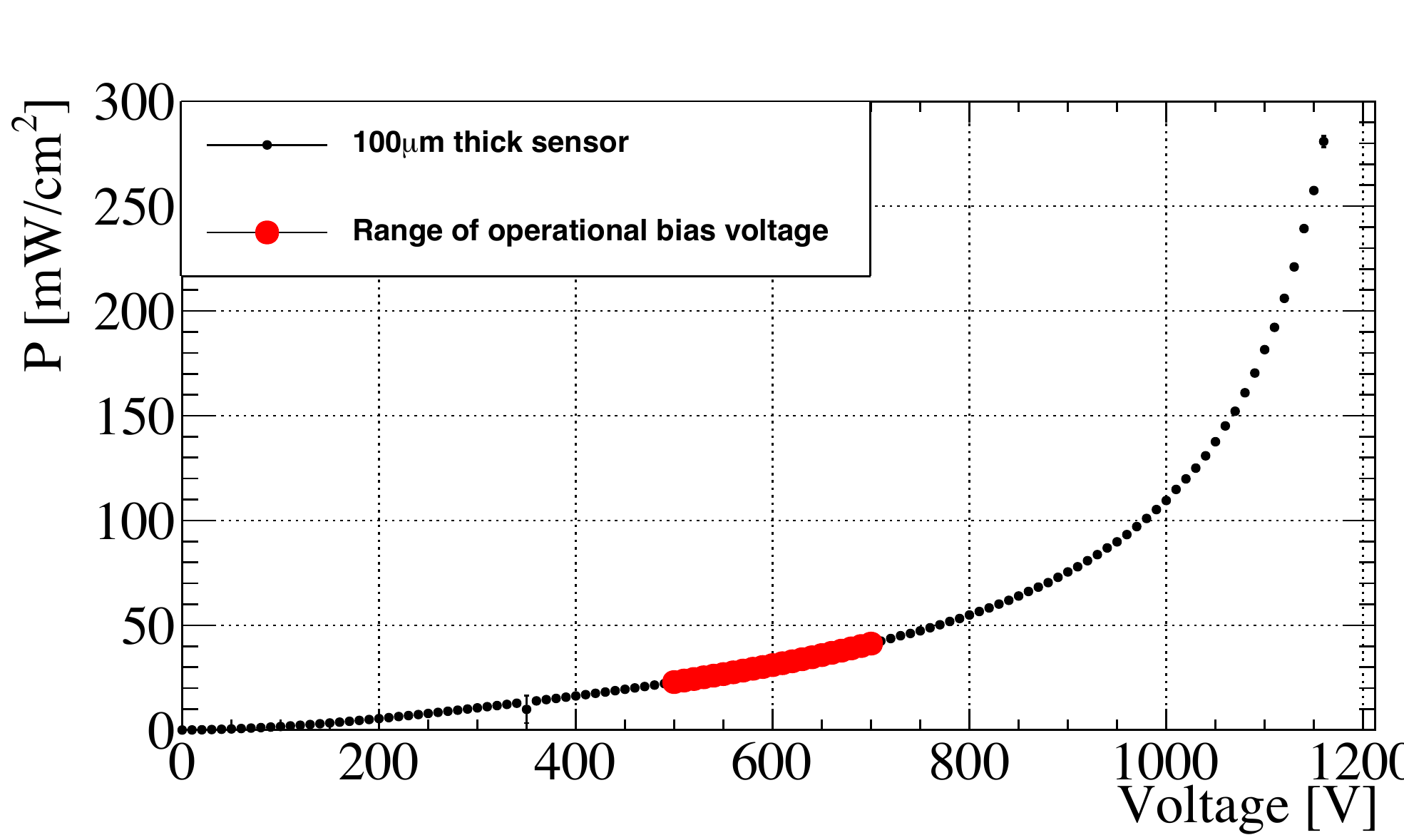}}
		\caption{(a) Comparison of hit efficiencies of FE-I4 modules irradiated from 2$\times$10$^{15}$ $\mathrm{n}_{\mathrm{eq}}/\mathrm{cm}^2$ to 1$\times$10$^{16}$ $\mathrm{n}_{\mathrm{eq}}/\mathrm{cm}^2$ at Josef-Stefan-Institute in Ljubljana, Slovenia and Karlsruhe Institute of Technology, Germany. (b) Power dissipation per area of a bare sensor at T=-25$^{\circ}$C irradiated to 1$\times$10$^{16}$ $\mathrm{n}_{\mathrm{eq}}/\mathrm{cm}^2$ and annealed for 11 days.}
		\label{vtt}
	\end{figure}

	An FE-I4 module from the VTT production with a 100~$\mathrm{\mu}$m thick sensor was irradiated to a fluence of 1$\times$10$^{16}$ $\mathrm{n}_{\mathrm{eq}}/\mathrm{cm}^2$ and measured in a beam test at DESY. In Fig.~\ref{vtt}(a) the hit efficiency is compared to previous measurements with other modules with 100~$\mathrm{\mu}$m thick sensors irradiated to lower fluences. The efficiencies at different fluences saturate at similar values with the main difference being the higher bias voltage needed to achieve the saturation for the module irradiated to 1$\times$10$^{16}$ $\mathrm{n}_{\mathrm{eq}}/\mathrm{cm}^2$. At the highest fluence a hit efficiency of 97.3\% is achieved at a moderate bias voltage of 500 V. It is thus possible to estimate a realistic range of operational bias voltage at this sensor thickness and fluence between 500 and 700 V including some safety factors. The efficiency loss is mainly caused by the standard punch-through design and, excluding the biasing structure in the calculation, results in an increase of the hit efficiency to a value of 98.4\%. Fig.~\ref{vtt}(b) shows the expected power density (power per sensor area) for the VTT module operated at a temperature of T=-25$^{\circ}$C as a function of the applied bias voltage up to 1200 V. To ensure a stable and reliably measured sensor temperature, the power consumption of bare sensors is measured on a cold chuck inside a probe station and is normalized to the active area of an FE-I4 module of 3.4~$\mathrm{cm}^{2}$. This procedure assures that the temperature of the sensor is known exactly thanks to the good thermal contact with the chuck. From this data, the resulting power density for a 100~$\mathrm{\mu}$m thick sensor operated at voltages of 500 to 700 V is determined to be 25-50~$\mathrm{mW}/\mathrm{cm}^2$. These low values make this technology an ideal candidate for use in the innermost layer for data taking at the HL-LHC.
	
	\section{Conclusions}
	Planar n-in-p pixel modules with and without active edge were investigated in view of the upgrade of the ATLAS pixel system for HL-LHC. The results of the characterization of FE-I4 modules employing a 100~$\mathrm{\mu}$m thick sensor with a 50~$\mathrm{\mu}$m active edge and one floating GR show a good performance with a hit efficiency of 99.8\% in the last pixel column and a mean hit efficiency of 60.4\% from the last pixel implant up to the sensor edge before irradiation. Modules with 100, 150 and 200~$\mathrm{\mu}$m thick sensors were analyzed regarding their cluster properties at high $\eta$ at HL-LHC for the future small pixel cell of 50x50~$\mathrm{\mu}$m$^{2}$. The thinner sensors were found to be favorable owing to their lower cluster size resulting in a reduced occupancy. Different targets of threshold in the range thought to be achievable with the future RD53A read-out chip were studied using a module employing a 150~$\mathrm{\mu}$m thick sensor. By decreasing the threshold from 1000 to 600~e the hit efficiency increased by 2\% up to 98\%. At CiS 50x250~$\mathrm{\mu}$m$^{2}$ pixel cells with 50x50~$\mathrm{\mu}$m$^{2}$ pixel implants were produced to investigate the performance of the future small pixel cells. The sensors are designed to be compatible with the current FE-I4 read-out chip. Two different sensor layouts to link five small implants to a 50x250~$\mathrm{\mu}$m$^{2}$ area using a metal rail were developed: adjacent pixels were linked one after each other forming a standard 50x250~$\mathrm{\mu}$m$^{2}$ pixel cell as well as linked in such a way that neighbouring pixel implants are read-out by two different channels allowing for an investigation of the effect of charge sharing in adjacent pixels. Both designs with a sensor thickness of 150~$\mathrm{\mu}$m show good performance before irradiation with comparable values of charge collection around 10~ke and hit efficiencies around 99\%. Measurements of modules with active edge sensors as well as modules with pixel cells with small 50x50~$\mathrm{\mu}$m$^{2}$ pixel implants are planned to be repeated after irradiation. Properties of an FE-I4 module with a 100~$\mathrm{\mu}$m thick sensor, irradiated at a fluence of 1$\times$10$^{16}$~\ensuremath{\mathrm{n}_{\mathrm{eq}}/\mathrm{cm}^2} and operated at bias voltages of 500 to 700 V are investigated. The module achieves a hit efficiency of 97.3\% at a power density of 25-50~\ensuremath{\mathrm{mW}/\mathrm{cm}^2} which makes this technology an ideal candidate for the innermost layer for data taking at HL-LHC.
	
	%\appendix
	%\section{Some title}
	
	\acknowledgments
	
	This work has been partially performed in the framework of the CERN RD50 Collaboration. The authors thank A. Dierlamm for the irradiation at KIT, V. Cindro and I. Mandic for the irradiation at JSI. Supported by the H2020 project AIDA-2020, GA no. 654168. 
	


\begin{thebibliography}{99}
		
		\bibitem{atlasUp} P.~S.~Miyagawa and I.~Dawson, {\it Radiation background studies for the Phase II inner tracker upgrade}, CERN, ATL-UPGRADE-PUB-2013-012 (2013).
		
		\bibitem{fei3} I.~Peric et al., {\it The FEI3 read out chip for the ATLAS pixel detector}, Nucl. Instr. Meth. A565 (2006) 178.
		
		\bibitem{fei4} M.~Garcia-Sciveres et al., {\it The FE-I4 pixel readout integrated circuit}, Nucl. Instr. Meth. A363 (2012) 29.
		
		\bibitem{rd53} RD53 Collaboration, \url{http://rd53.web.cern.ch/RD53/}
		
		\bibitem{chip} M.~Garcia-Sciveres et al., {\it Towards third generation pixel readout chips}, Nucl. Instr. Meth. A731 (2013) 83-87.
		
		\bibitem{usbpix} USB based readout system for ATLAS FE-I3 and FE-I4, \url{http://icwiki.physik.uni-bonn.de/twiki/bin/view/Systems/UsbPix}
		
		\bibitem{rce} RCE Distribution for ATLAS Applications - SLAC, \url{http://www.slac.stanford.edu/exp/atlas/upgrade/RCE-distribution-2015Mar.html}
		
		\bibitem{jens} J.~Weingarten et al., {\it Planar Pixel Sensors for the ATLAS Upgrade: Beam Tests results}, JINST 7 (2012) P10028. 
		
		\bibitem{hendrick} H.~Jansen et al.,  {\it Performance of the EUDET-type beam telescopes}, arXiv:1603.09669, DESY-16-055, (2016).
		
		\bibitem{eutelescope} H.~Perrey, {\it EUDAQ and EUTelescope Software Frameworks for Testbeam Data Acquisition and Analysis}, Proceedings for Tipp2014, Amsterdam, The Netherlands, PoS, (2014).
		
		\bibitem{phil} P.~Weigell, {\it Investigations of Properties of Novel Silicon Pixel Assemblies Employing Thin n-in-p Sensors and 3D-Integration}, PhD thesis, MPP-2013-5 (2013).
		
		\bibitem{edge} X.~Wu et al., {\it Recent advances in processing and characterization of edgeless detectors}, JINST 7 (2012) C02001.
		
		\bibitem{nani} N.~Savic, {\it Thin n-in-p planar pixel modules for the ATLAS upgrade at HL-LHC}, Proceedings of the Vienna Conference on Instrumentation 2016, Nucl. Instr. Meth. A (2016).
		
		\bibitem{wittig} T.~Wittig, {\it Ongoing activities at CiS}, RD50 talk at 27th RD50 Conference at CERN, \url{https://indico.cern.ch/event/456679/}, (2015).
		
		\bibitem{anna} A.~Macchiolo et al., {\it Development of n-in-p pixel modules for the ATLAS upgrade at HL-LHC}, Proceedings of HSTD10 , Xian, China, Nucl. Instr. Meth. A (2016).
		
		
		% Please avoid comments such as "For a review'', "For some examples",
		% "and references therein" or move them in the text. In general,
		% please leave only references in the bibliography and move all
		% accessory text in footnotes.
		
		% Also, please have only one work for each \bibitem.
		
		% Please avoid comments such as "For a review'', "For some examples",
		% "and references therein" or move them in the text. In general,
		% please leave only references in the bibliography and move all
		% accessory text in footnotes.
		
		% Also, please have only one work for each \bibitem.
		
		
	\end{thebibliography}
\end{document}